\documentclass[doublecol, a4paper]{epl2} 

\usepackage{psfrag}
\usepackage{graphicx}
\usepackage{color}

\usepackage[latin1]{inputenc}
\usepackage{graphicx}
\usepackage{amsmath}
\usepackage{amssymb}
\usepackage{relsize}

\newcommand{\pd}[2]{\frac{\partial #1}{\partial #2}}
\newcommand{\mean}[1]{\left \langle #1 \right \rangle}

\newcommand{\lam}{\lambda}
\newcommand{\beq}{\begin{equation}}
\newcommand{\eeq}{\end{equation}}
\newcommand{\al} {\alpha}

\newcommand{\nn} {\nonumber}

\newcommand{\xvar} {w}

\newcommand{\Teff} {T_{\rm{eff}}}

\newcommand{\etaqs} {\eta_{\rm{qs}}}
\newcommand{\etac} {\eta_{\rm{C}}}
\newcommand{\etaca} {\eta_{\rm{CA}}}

\newcommand{\Th} {T_{\rm{h}}}
\newcommand{\Tc} {T_{\rm{c}}}

\newcommand{\Airrf}[1] {A_{\rm{irr}}^{(#1)}}
\newcommand{\Airr} {A_{\rm{irr}}}

\newcommand{\Wirr} {W_{\rm{irr}}}

\newcommand{\x} {\mathbf{x}}

\newcommand{\tf} {t_1}
\newcommand{\tb} {t_3}

\newcommand{\ttot} {t_{\rm{tot}}}

\newcommand{\jb} {\mathbf{j}}

\newcommand{\wa} {w_a}
\newcommand{\wb} {w_b}
\newcommand{\pa} {p_a}
\newcommand{\pb} {p_b}

\newcommand{\pn} {\mathcal{P}}
\newcommand{\jbn} {\mathcal{J}}

\newcommand{\tges} {(t_f - t_i)}
\newcommand{\tauel} {(\tau - t_i)}
\newcommand{\taun} {\tauel / \tges}

\begin{document}

	\title{Efficiency at maximum power: An analytically solvable model for stochastic heat engines}
\author{Tim Schmiedl and Udo Seifert}

\institute{                    
  {II.} Institut f\"ur Theoretische Physik, Universit\"at Stuttgart,
  70550 Stuttgart, Germany
}

\begin{abstract}
{We study a class of cyclic Brownian heat engines in the framework of finite-time thermodynamics. For infinitely long cycle times, the engine works at the Carnot efficiency limit producing, however, zero power. For the efficiency at maximum power, we find a universal expression, different from the endoreversible Curzon-Ahlborn efficiency. Our results are illustrated with a simple one-dimensional engine working in and with a time-dependent harmonic potential.}
\end{abstract}

 {Brownian Motion}
 {Nonequilibrium and irreversible thermodynamics}
 {Colloids}
 {Fluctuations}

\def\la{\lambda(\tau)}
\def\ls{\lambda^*(\tau)}
\def\li{\lambda_i}
\def\wmus{W[\la]}

\maketitle
\section{Introduction}

According to Carnot, the second law leads to a maximal efficiency $\etac \equiv 1 - \Tc / \Th$ for heat engines working between two heat baths at temperatures $\Th > \Tc$. However, this efficiency can only be achieved in the quasistatic limit, where transitions occur infinitesimally slowly and hence the power output vanishes.
For real heat engines, it is more meaningful to calculate the efficiency at the maximal power output of the machine. The Curzon-Ahlborn efficiency at maximum power $\etaca \equiv 1 - \sqrt{\Tc / \Th}$  was originally derived for a Carnot-engine with finite thermal resistance coupling to the thermal reservoirs \cite{curz75}. This behaviour is even recovered if path optimization techniques are used to maximize the work output with respect to the driving scheme, see \cite{hoff03} and references therein.  Recent extensions \cite{vdb05, cisn07} of this apparent universal law to an infinitesimal series of coupled heat engines operating in a linear response regime (i.e. between heat baths with small temperature difference) have again risen the question whether the Curzon-Ahlborn result is quite generally a bound on the efficiency of heat engines working at maximum power output. 

In contrast to the macroscopic heat engines considered in conventional endoreversible thermodynamics, thermal fluctuations play a crucial role in most biologically relevant systems and dynamics cannot be described on a deterministic (macroscopic) level. In this regime, models for thermodynamic machines must incorporate fluctuation effects and thus allow also for backward steps even in a directed motion \cite{juel97, astu02}. Brownian motors, in contrast to genuine heat engines, are mostly driven by time-dependent potentials or chemical potential differences. In the last two decades, a variety of aspects of such Brownian motors has been studied, including dynamics \cite{astu02, reim02a}, stochastic energetics \cite{seki97, seki98, seki00, parr02} and efficiencies \cite{parm99, hond00, vdb07}, both on a continuous (Langevin equation) and a discrete (Master equation) state space. 
Since biological motors are typically driven far out of equilibrium, linear response thermodynamics is not appropriate to describe these systems. The seminal work of Sekimoto \cite{seki97,seki98} opened the door for a thermodynamic description of Langevin systems driven far out of equilibrium. Thermodynamic quantities such as work, heat, internal energy, and entropy can even be defined on a single stochastic trajectory \cite{seki98, seif05a, schm06a}, yielding the respective ensemble quantities after averaging. 

Brownian heat engines have been investigated with a model system using a spatial temperature profile and a ratchet potential \cite{feynman, buet87, land88}. This thermal ratchet model has also been extended to a Brownian particle in the underdamped regime \cite{blan98}.  Efficiencies of such ratchet heat engines \cite{seki97, jarz99a, asfa04, gome06} and related Brownian heat engines \cite{vdb04} have been calculated. Recent studies on the efficiency at maximum power, however, have either been restricted to the linear response regime \cite{gome06} or use a questionable definition of work (which allows for the efficiency to exceed the Carnot bound) \cite{asfa04}. For a ratchet heat engine on a discrete state space, a violation of the Curzon-Ahlborn bound has been found \cite{vela01}. 

Although widely used as a model system for a stochastic heat engine, ratchet heat engines are hard to realize experimentally because of the necessary temperature gradients on a small length scale. Moreover, a macroscopic limit of such heat engines does not lead to heat engines of the Carnot type. In order to overcome these difficulties, we introduce a simple model system of a stochastic heat engine whose degrees of freedom are subject to a time-dependent potential with time-dependent coupling to and decoupling from the two heat baths. This model is somewhat distinct from the thermal ratchets which are driven by spatial temperature differences. Ratchet heat engines with a time-dependent but spatially constant temperature have been studied in \cite{reim96, soko97, kuli05}.

We first describe the general setup of a stochastic heat engine consisting of two adiabatic and two isothermal steps. We then illustrate this setup by a simple one-dimensinal example and show that our model exhibits a universal dependence of work and heat on the cycling time. This leads to a universal form of the efficiency at maximum power. We first neglect the effect of kinetic energy in our overdamped description and discuss its role (cf. \cite{hond00}) at the end.

\section{The model}
We consider a thermodynamic machine with internal degrees of freedom ${\bf x}$ and an external time-dependent potential $V({\bf x},\tau)$. On a microscopic scale, it is crucial to include thermal fluctuations in the description and thus to consider the probability density $p(\x, \tau)$ to find the system in state $\x$ at time $\tau$. In the overdamped limit, the time evolution of  $p(\x,\tau)$ is governed by the Fokker-Planck equation
\beq
\partial_\tau p(\x,\tau) = -\nabla \cdot \jb = -\nabla \cdot {\bf \mu}  \left[- \left ( \nabla V \right )  -  k_B T \nabla \right] p(\x,\tau)
\label{FP}
\eeq
where ${\bf \mu}$ is the mobility matrix. Similar to a macroscopic Carnot engine, we consider the machine cycle to consist of two isothermal and two adiabatic transitions.

{\sl Isothermal steps .--}
The potential $V(\x,\tau)$ is varied during a time-interval $t_i\leq \tau\leq t_f$ at a given constant temperature $T$.
The mean work $W$ spent in this process (or extracted for $W < 0$)
\begin{eqnarray}
W &=& \int_{t_i}^{t_f} d\tau \int dx \ p(\x,\tau)  \pd V \tau \nn \\
&=& \Wirr - T \Delta S + \Delta E 
\label{meanw}
\end{eqnarray}
becomes a functional of the time dependent potential $V(\x, \tau)$. For notational simplicity, we set $k_B=1$ in the following. In Eq. (\ref{meanw}), we have introduced the (mean) irreversible work 
\beq
\Wirr \equiv   \int_{t_i}^{t_f} d\tau \int dx \   \frac {\jb(\x,\tau) {\bf \mu}^{-1} \jb(\x,\tau)} {p(\x,\tau)},
\label{W_irr}
\eeq
the (mean) internal energy 
\beq
E(\tau) \equiv \int dx \  p(\x,\tau) V(\x,\tau),
\eeq
and the system entropy 
\beq
S(\tau) \equiv -\int dx \ p(\x,\tau) \ln (p(\x,\tau)),
\eeq
see also \cite{seif05a, schm06a}.  

{\sl Adiabatic steps .--}
Adiabatic steps are somewhat more difficult to conceive realistically (just as for macroscopic heat engines). Here, they are idealized as sudden jumps of the potential while uncoupling from one heat bath and coupling to the other heat bath occurs instantaneously and thus without heat exchange. Note that, within our overdamped description, we first neglect kinetic energies. Considering the kinetic energy of the degrees of freedom would lead to an additional heat flux in these steps because of the relaxation of the momentum due to the temperature change. This issue is discussed in detail at the end of this Letter. Since there is no time for relaxation, the distribution $p(\x,\tau)$ and thus the system entropy $S$ does not change during these steps and hence these transitions are also isentropic. The mean work applied in such a transition occuring at time $\tau$ is given by
\beq
W = \Delta E = \int dx \  p(\x,\tau) \left[ V(\x, \tau + 0) - V(\x, \tau-0) \right ].
\eeq

\begin {figure}
\begin{center}
\includegraphics[width = 0.47 \textwidth]{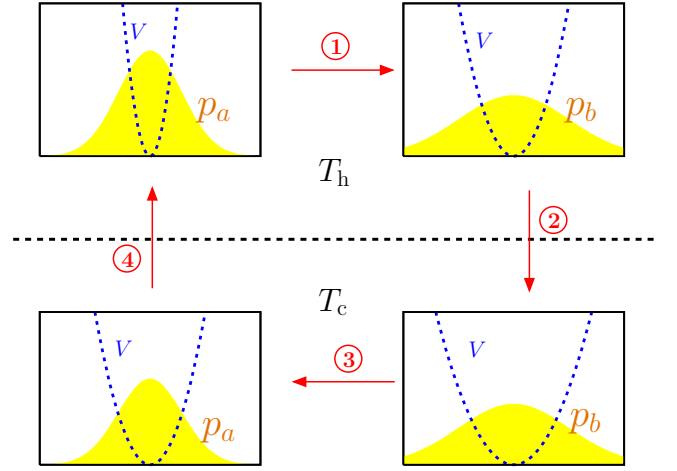} 
\caption{Scheme for a cyclic process of a stochastic heat engine operating between heat baths at temperature $\Th$ and $\Tc$. The dotted curve shows a one-dimensional potential $V(x,\tau)$ and the filled curve the boundary values $\pa(x)$, $\pb(x)$ of the corresponding distribution $p(x,\tau)$. Steps 1 and 3 are isothermal whereas steps 2 and 4 are adiabatic.  
\label{qs_scheme}}
\end{center}
\end {figure}

{\sl Cyclic process.--}
We construct a cyclic process by performing subsequently the following four steps, see Fig. \ref{qs_scheme}.

(1) Isothermal transition at temperature $\Th$ during time interval $0 < \tau < \tf$. The potential $V(\x,\tau)$ is changed time-dependently and work is extracted ($W < 0$).

(2) Adiabatic transition (instantaneously) from temperature $\Th$ to temperature $\Tc$ at time $\tau = \tf$. The distribution $\pb(\x) \equiv p(\x, \tau = \tf)$ does not change during this step.

(3) Isothermal transition at temperature $\Tc$ during time interval $\tf < \tau < \tf + \tb$. The potential $V(\x,\tau)$ is changed time-dependently (with $W > 0$).

(4) Adiabatic transition from temperature $\Tc$ to temperature $\Th$ at time $\tau = \tf + \tb$. The distribution $\pa(\x) \equiv p(\x, \tau = \tf + \tb)$ does not change during this step.

For a cyclic process, periodicity in time requires $p(\x, \tf+\tb) = p(\x, 0)$ thus imposing a constraint on the admissible time-dependence of $V(\x,\tau)$. Moreover, a cyclic process implies $\sum_{i=1}^4 \Delta E^{(i)} = \sum_{i=1}^4 \Delta S^{(i)} = 0$. Since the system entropy $S$ does not change during the adiabatic steps, $\Delta S^{(2)} = \Delta S^{(4)} = 0$, we even have $\Delta S^{(1)} = -\Delta S^{(3)} \equiv \Delta S$.
The total work then is given by
\begin{eqnarray}
W &=& \sum_{i=1}^4 W^{(i)} = \sum_{i=1}^4 \Wirr^{(i)}  -\sum_{i=1}^4 T_i \Delta S^{(i)}  + \sum_{i=1}^4 \Delta E^{(i)} \nn \\
  &=& \Wirr^{(1)} + \Wirr^{(3)} - (\Th - \Tc) \Delta S.
\label{Wges}
\end{eqnarray}
The heat uptake during one complete cycle from the hotter heat bath at temperature $\Th$ is
\begin{eqnarray}
-Q^{(1)} &\equiv& -W^{(1)} + \Delta E^{(1)} = \Th \Delta S - \Wirr^{(1)}.
\label{Q_in_gen}
\end{eqnarray}

\section{Example}
In order to illustrate this general concept of a stochastic heat engine, we consider
the motion of a colloidal particle in a trap with time-dependent strength. The corresponding
potential reads 
\beq
V(x,\tau) =  {\lam(\tau)} x^2 / {2} .
\eeq
The probability distribution $p(x,\tau)$ remains Gaussian for all times $\tau$ if it is so initially. We first derive the equation of motion for the variance $\xvar(\tau) \equiv \mean{x^2(\tau)}$
\beq
\dot \xvar = -2 \mu \lam \xvar + 2 \mu T
\label{dvar}
\eeq
by multiplying eq. (\ref{FP}) with $x^2$ and integrating over $x$, where we denote time-derivatives by a dot.
For any initial Gaussian distribution $p(x,0)$, the internal energy of the system is given as
\beq
E(\tau) = \int dx \ p(x,\tau) V(x) =  \xvar(\tau) \lambda(\tau) / 2 .
\eeq
Using (\ref{dvar}), the mean work (\ref{meanw}) can be cast in a local functional of the new variable $\xvar$ and its first derivative by solving (\ref{dvar}) for $\lam$. For a constant temperature $T$, this yields
\begin{eqnarray}
W[\lam(\tau)] &=& \int_{t_i}^{t_f} d\tau \dot \lam \frac {\xvar} 2 = \Wirr - T \Delta S + \Delta E \label{W_wang} \\
&=&  \frac {1} {4 \mu} \int_{t_i}^{t_f} d\tau \frac {{\dot \xvar}^2} {\xvar} - \frac 1 2 T \left [\ln \xvar \right]_{t_i}^{t_f} + \frac 1 2 \left [ \lam \xvar \right]_{t_i}^{t_f} . \nn
\end{eqnarray}

\begin {figure}
\begin{center}
\includegraphics[width = 0.49\textwidth]{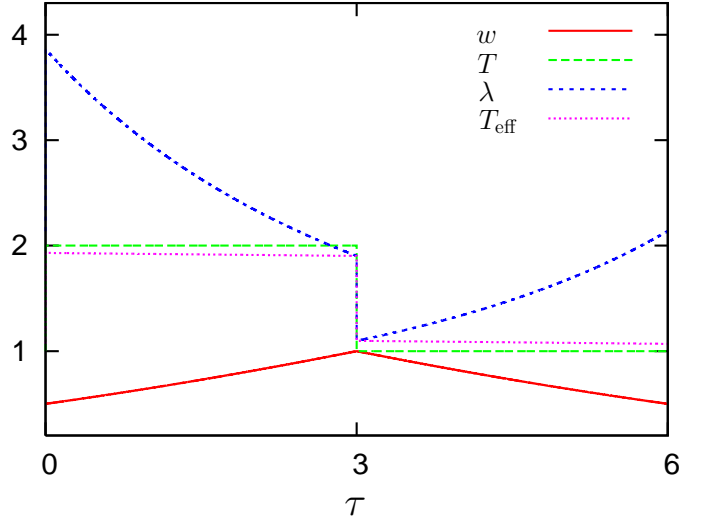}
\caption{
Stochastic heat engine in a harmonic potential: Time dependence of the variance of the particle position $\xvar$, the temperature $T$, the trap strength $\lam$ and the effective temperature $\Teff$ (Eq. (\ref{Teff})) for one cycle with $\tf = \tb = 3$, $\wa = 0.5$, $\wb = 1$, $T_h = 2$, $T_c = 1$, and $\mu = 1$, starting with step $1$. 
\label{finite-time}}
\end{center}
\end {figure}

We now have to choose a protocol $\lambda(\tau)$ for the isothermal transitions.  The variance $\xvar$ of the particle position depends on this protocol via eq. (\ref{dvar}). As a consequence, the mean work (\ref{W_wang}) will also depend on this protocol. We ask for the optimal protocols $\lam_1^*(\tau)$ and $\lam_3^*(\tau)$ yielding a minimal total work (equivalent to a maximal work output). Rather than imposing boundary conditions on the control parameter $\lam$, we impose boundary values of the particle position variances $\wa \equiv w(0) = w(\tf + \tb)$ and $\wb \equiv w(\tf)$ for a given finite cycle time $\ttot \equiv \tf + \tb$. While it might seem more natural to impose a range of (experimentally) possible $\lam$ values, such an approach can lead to unphysical optimal processes which are driven infinitely fast with infinitesimally small changes in the variance $\xvar(\tau)$ (data not shown). With slight modification of the results in \cite{schm07} due to the different boundary conditions, we get the optimal protocols for the variances
\begin {eqnarray}
\xvar_1^*(\tau) = \wa \left (1 + (\sqrt{\wb / \wa} - 1) \tau / \tf \right )^2 \\
\xvar_3^*(\tau) = \wb \left (1 + (\sqrt{\wa / \wb} - 1) (\tau-\tf) / \tb \right )^2.
\end{eqnarray}
The optimal protocols $\lam_i^*(\tau)$ for the strength of the trap can then be calculated from Eq. (\ref{dvar}). The resulting quantities are shown in Fig. \ref{finite-time}.

For this cyclic process, we get the total work output
\begin{eqnarray}
-W &=&  \frac 1 2 (\Th - \Tc) \ln \frac {\wb} {\wa} - \frac 1 \mu \left ( \frac 1 {\tf} + \frac 1 {\tb} \right ) {(\sqrt{\wb} - \sqrt{\wa})^2}  \nn \\ 
&\equiv& (\Th - \Tc) \Delta S - \left ( \frac {1} {\tf} + \frac {1} {\tb} \right ) \Airr ,
\label{W_eng}
\end{eqnarray}
with the irreversible ``action'' $\Airr$. The work only depends on the ``reduced time'' $t_r \equiv 1/ (1/\tf + 1/\tb) = \tf \tb / (\tf + \tb)$. 

The heat uptake during one complete cycle from the hotter heat bath at temperature $\Th$ is given as 
\begin{eqnarray}
-Q^{(1)} = \Th \Delta S - \Wirr^{(1)} =  \Th \Delta S - \frac {\Airr}{\tf} 
\label{Q_in}
\end{eqnarray}
yielding the efficiency
\beq
\eta \equiv \frac {-W}{-Q^{(1)}} = \frac {(\Th - \Tc) \Delta S - {\Airr / t_r}} {T_h \Delta S - {\Airr / \tf}}
\eeq

For $\tf, \tb \to \infty$, we recover the Carnot efficiency $\eta_c \equiv 1- \Tc / \Th$. This shows that the sudden adiabatic steps do not spoil the reversibility.

We immediately recognize from Eq. (\ref{W_eng}) that there exists a ``stall time'' 
\beq
t_r^{\rm{st}} \equiv \frac {\Airr} {(\Th - \Tc) \Delta S} = \frac  { 2 \left (  \sqrt{\wb} - \sqrt{\wa} \right)^2 } {\mu (\Th - \Tc) \ln \left ( \wb / \wa \right) }
\eeq
where the heat engine does not perform work anymore. Rather, at smaller times $t_r < t_r^{\rm{st}}$, it consumes work.

We next explore how fast the heat engine must be driven to yield a maximal power output. The power $P = -W / (\tf + \tb)$ has a maximum for
\beq
\tf^{*} = \tb^{*} =  \frac {4 \Airr} {(\Th - \Tc) \Delta S}
\eeq
which yields a reduced optimal time 
\beq
t_r^{*} = \frac 1 2 \tf^{*} = 2 t_r^{\rm{st}}. 
\eeq
A similar relation between the stall force and the force at maximum power has been obtained in \cite{vdb05} within a linear response treatment.
The total work output at maximum power
\beq
-W^{*} =  \frac 1 2 (\Th - \Tc) \Delta S 
\eeq
and the corresponding heat uptake
\beq
-Q^{* (1)} = \frac 1 4  \Delta S (3 \Th + \Tc) 
\label{Q_in_max}
\eeq
then yield the efficiency at maximum power
\beq
\eta^{*} = \frac {2 (\Th - \Tc)} {3 \Th + \Tc} = \frac {\etac}{2-\etac / 2} .
\label{eta_max}
\eeq
This result is independent of the imposed boundary values $\wa$ and $\wb$, suggesting a certain degree of universality confirmed below. For large temperature differences, $\Th - \Tc \gg \Tc$, the efficiency has the limit $\eta^{*} \to 2 / 3$.
An expansion for small relative temperature differences, $\mathcal{T} \equiv (\Th-\Tc)/\Tc \ll 1$,
\beq
\eta^{*} = \frac {2 \mathcal{T}} {3 \mathcal{T} + 4} = \frac {\mathcal{T}} 2 - \frac {3 \mathcal{T}^2}{8} + \frac {9 \mathcal{T}^3}{32} + \mathcal{O} \left ( {\mathcal{T}^4} \right)
\eeq
shows that the deviation from the Curzon-Ahlborn efficiency 
\beq
\eta_{\rm{CA}} = 1 - \sqrt{\Tc/\Th} =  \frac {\mathcal{T}} 2 - \frac {3 \mathcal{T}^2}{8} + \frac {10 \mathcal{T}^3}{32} + \mathcal{O} \left ( {\mathcal{T}^4} \right)
\eeq
occurs at the third order in the relative temperature difference $\mathcal{T}$. The validity of the Curzon-Ahlborn efficiency in a linear response regime has been shown quite generally in \cite{vdb05}. We recover this result even including the second order in the relative temperature difference, which goes beyond linear response. For larger deviations from equilibrium, however, the efficiency at maximum power in this example is smaller than the Curzon-Ahlborn bound.

\label{illus}

\section{General Case}

We now relax the restriction to one degree of freedom in a harmonic potential and go back to the general case where we also impose two boundary distributions: $\pa(\x)$ at $\tau = 0$ (and at $\tau = \tf + \tb$) and $\pb(\x)$ at $\tau = \tf$ for the two isothermal transitions. 
As seen from Eq. (\ref{Wges}), minimizing the total mean work $W$ (leading to a maximal work output) subject to such boundary conditions, which leave the $\Delta S$ term invariant, is equivalent to the minimization of the irreversible work (\ref{W_irr}) for both isothermal transitions. Using a normalized time $\hat \tau \equiv \taun$ and thus a time-dependent distribution $\mathcal{P} (\x, \hat \tau)$, it follows from (\ref{FP}) that the current can be expressed as $\jb = \mathcal{J}(\x, \hat \tau) / \tges$. The irreversible work (\ref{W_irr}) can then be written as 
\beq
\Wirr \equiv \frac {1} {\tges} \int_{0}^{1} d \hat \tau \int dx \   \frac {\jbn(\x,\hat \tau) {\bf \mu}^{-1} \jbn(\x,\hat \tau)} {\pn(\x,\hat \tau )}.
\label{W_irr_2}
\eeq
Minimizing the irreversible work $\Wirr$ thus is equivalent to minimizing the integral in (\ref{W_irr_2}) which only depends on $\pn(\x, \hat \tau)$ and $\jbn(\x, \hat \tau)$. The time evolution $\partial_\tau p = - \nabla \cdot {\bf j}$ expressed in the time-normalized quantities imposes the constraint $\partial_{\hat \tau} \pn = - \nabla \cdot {\bf \jbn}$. Minimizing the integral in (\ref{W_irr_2}) subject to this constraint leads to the optimal distribution $\pn^*(\x, \hat \tau)$ which does not depend on the total transition time. Since this optimization has to be done for each of the isothermal transitions separately, we get optimal distributions $\pn_1^*(\x,\tau /t_1)$ and $\pn_3^*(\x, (\tau - \tf)/t_3)$ for the two isothermal transitions which both are independent of the total transition times $t_j$, $j = 1,3$. 

The optimization with respect to maximal work output is tractable analytically only if we restrict to harmonic potentials $V(\x,\tau)$, as shown above in the example. Otherwise, the resulting Euler-Lagrange equations are non-linear partial differential equations and even the numerical solution seems to be a difficult task.
\begin {figure}
\begin{center}
\includegraphics[width = 0.49\textwidth]{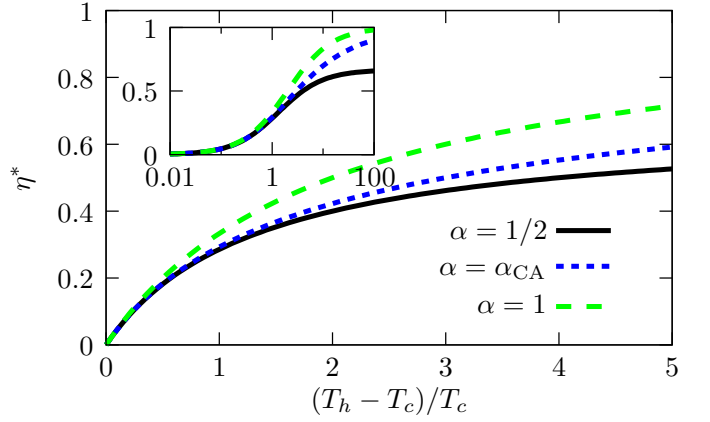}
\caption{
Efficiency $\eta^*$ at maximum power as a function of $(\Th-\Tc)/\Tc$ for $\al = 1/2$, $\al_{\rm{CA}}$, and $1$, corresponding to the temperature independent mobility of the example, the Curzon-Ahlborn efficiency, and the upper bound of (\ref{eta_max_gen}), respectively. The inset shows the same data in a log-plot. \label{fig_eta_gen}}
\end{center}
\end {figure}
For a derivation of our central result, however, we do not need an explicit solution. It suffices to observe that Eq. (\ref{W_irr_2}) implies a remarkable universal feature for the optimal driving scheme: The (mean) irreversible work in the isothermal transitions scales exactly with the inverse transition time 
\beq
\Wirr^{(j)} =  \Airrf{j} / t_j,
\eeq
where $\Airrf{j}$ is independent of the transition times $t_j (j = 1, 3)$ and becomes a functional only of $\pa(\x)$, $\pb(\x)$, and the mobility matrix $\mu$.
As a direct consequence, the work output during a complete cycle is given by
\beq
-W = (\Th - \Tc) \Delta S -  \frac {\Airrf{1}} {\tf} - \frac {\Airrf{3}} {\tb} ,
\label{W_tot_gen}
\eeq
The heat flux from the hotter bath $\Th$ is
\beq
-Q^{(1)} = \Th \Delta S - \frac {\Airrf{1}} {\tf}.
\label{Q_tot_gen}
\eeq

For negative $\Delta S < 0$ and transition times larger than the ``stall time'' $t > t^{\rm{st}}$, where the engine does not perform work anymore, the setup performs as a heat engine with efficiency 
\beq
\eta \equiv \frac{-W} {-Q^{(1)}}
\eeq
approaching the Carnot efficiency $\etac \equiv 1 - \Tc / \Th$ for infinite cycle time $t \to \infty$. 
Maximization of $P \equiv -W / (\tf + \tb)$ with respect to the transition times $\tf, \tb$ leads to a universal expression for the efficiency at maximum power of
\beq
\eta^*(\al) = \frac {\etac} {2- \al \etac}
\label{eta_max_gen}
\eeq
where 
\beq
\al \equiv  1  / \left ( { 1 + \sqrt{{\Airrf{3} / {\Airrf{1}}}}} \right ) .
\eeq
This expression for the efficiency at maximal power is our main result. In the linear response limit, $\Th - \Tc \ll \Tc$, we recover $\eta^* \to \etac / 2$, consistent with the analysis in \cite{vdb05}. Beyond linear response, one needs the irreversible action $\Airrf{i}$ which, in principle, can be calculated for any specific system by solving the optimization problem. In general, due to the temperature dependence of the mobility $\mu$, the irreversible action $\Airrf{i}$ is temperature dependent. An explicit statement becomes possible for an isotropic mobility matrix $\mu$ in Eq. (\ref{FP}). In this case, $\al$ depends only on the temperature dependence of $\mu$ via ${\Airrf{3} / \Airrf{1}} =  \mu(\Th) / \mu(\Tc)$. For a temperature independent mobility, i. e. $\al = 1 / 2$, we recover the result (\ref{eta_max}) from the example above which thus holds also in higher dimensions.
Only for $\mu \sim 1/T$, corresponding to $\al = \al_{\rm{CA}} \equiv 1 / (1+\sqrt{\Tc / \Th})$, we recover the Curzon-Ahlborn efficiency $\eta^* = 1 - \sqrt{\Tc / \Th}$.  These efficiencies at maximum power are shown in Fig. \ref{fig_eta_gen}. 
From a theoretical perspective, it is important to note that for $1 > \al > \al_{\rm{CA}}$, the result (\ref{eta_max_gen}) exceeds the Curzon-Ahlborn bound. While in the present setup this can be achieved only if the mobility $\mu$ decreases faster than $\sim 1 / T$ with temperature, $\al = 1$ can be reached in discrete ratchet heat engines \cite{vela01}.

As outlined in the introduction, the Curzon-Ahlborn efficiency is a rather universal bound on the efficiency at maximum power for a wide class of model heat engines. Why does the efficiency at maximum power derived here also under quite general conditions differ significantly from the Curzon-Ahlborn result?

A first hint is given by analyzing our example in analogy to an endoreversible Curzon-Ahlborn machine. For the latter, a linear Fourier law for the heat flow is assumed. However, for systems driven strongly out of equilibrium, non-linearities can occur. In order to analyze our model heat engine in this framework, we first have to assign an ``effective'' temperature to the working medium (the Brownian particle). Since in this harmonic potential the distribution remains equilibrium-like (Gaussian) even out of equilibrium, we can assign an effective temperature
\beq
\Teff(\tau) \equiv  {\xvar(\tau)}{ \lambda(\tau)}
\label{Teff}
\eeq
to the system. If we (instantaneously) coupled a heat bath at this temperature $\Teff$ to our system and kept the current trap strength $\lambda$ constant, we would recover an equilibrium situation.
We now express the heat exchange of the system in terms of this effective temperature
\begin{eqnarray}
dQ &=& dW - dE = \frac 1 2 \xvar d \lambda - \frac 1 2 d (\xvar \lambda) = -\frac 1 2 \lam d \xvar \nn \\
 &=& - \frac 1 2 \lam \dot \xvar d \tau = - \mu \lam (T - \lam \xvar) d \tau \nn \\
&=& - \mu \lam(\tau) (T-\Teff) d\tau
\end{eqnarray}
and recover a linear Fourier law for the heat flow. However, the thermal conductivity $\mu \lam(\tau)$ becomes time-dependent in contrast to the assumptions of the endoreversible Curzon-Ahlborn engine.

Secondly, the derivation of an apparent universal Curzon-Ahlborn bound in Refs. \cite{vdb05, cisn07} by using a cascade construction of $N$ machines working in the linear response regime comes at the price of an infinitesimally small power output in the (implicitly assumed) limit $N \to \infty$. For the setup in \cite{vdb05}, it can easily been shown that for a mechanical heat engine in the linear response regime, the maximal power output scales as $P  \sim (\Delta T)^2$ where $\Delta T \equiv \Th - \Tc$. The coupling of $N$ such machines with equal temperature differences $\Delta T = (\Th - \Tc) / N$ thus leads to a total maximal power output of $P^*  \sim 1 / N$.

\section{Role of kinetic energy}

So far, we have neglected the kinetic energy of the degrees of freedom. However, even in the overdamped limit, coupling to and decoupling from the heat baths leads to a heat flux due to the kinetic energy of the particle \cite{hond00}. This additional heat flux $Q_{\rm{kin}} = (\Th - \Tc) / 2$ per degree of freedom can easily be considered, leading to a general decrease of efficiencies. For large changes of the potential $V(\x,\tau)$ during the cycle, however, this contribution is small and can be neglected. 
In general, a given amount of heat leakage $Q^{\rm{leak}}$ per cycle (e. g., due to the kinetic energy) leads to an additional term in Eq. (\ref{Q_tot_gen})
\beq
-Q^{(1)} = \Th \Delta S + Q^{\rm{leak}} - \frac {\Airrf{1}} {\tf} .
\label{Q_tot_gen_leak}
\eeq
For $Q^{\rm{leak}}>0$, Carnot efficiency cannot be reached anymore and for $\tf \to \infty$, $\tb \to \infty$, we get the quasistatic efficiency 
\beq
\eta_{\rm{qs}} \equiv  (\Th - \Tc) \Delta S / (\Th \Delta S + Q^{\rm{leak}}).
\eeq
A straightforward calculation shows that the result (\ref{eta_max_gen}) then is generalized to
\beq
\eta^*(r) = \frac {\etaqs} {2- \al \etaqs}
\label{eta_max_gen_qs}
\eeq
where $\etac$ now is replaced by the quasistatic efficiency $\etaqs$.

\section {Summary}

In summary, we have analyzed the efficiency at maximal power for a large class of stochastic heat engines. We find a universal law (\ref{eta_max_gen}) differing substantially from the prevailing Curzon-Ahlborn bound. Even though we here derived the result using the concept of stochastic thermodynamics, one might expect that under appropriate conditions a ``thermodynamic limit'' exists, bridging the gap between the mesoscopic system discussed here and macroscopic heat engines.

\end{document}